\documentclass[aps,prl,twocolumn,showpacs,superscriptaddress,groupedaddress]{revtex4}  % for review and submission
\usepackage{graphicx}  % needed for figures
\usepackage{dcolumn}   % needed for some tables
\usepackage{bm}        % for math
\usepackage{amssymb}   % for math
\usepackage{natbib}
\setcitestyle{square, comma, numbers,sort&compress}

\begin{document}

\widetext
\leftline{Version xx as of \today}
\leftline{Primary authors: Petr B\'{\i}lek, J\'an Tungli, Milan \v{S}imek, Zden\v ek Bonaventura}
\leftline{To be submitted to PSST}
\author{Petr B\'{\i}lek}
\affiliation{Department of Physical Electronics, Faculty of Science, Masaryk University, Brno, Czech Republic}
\affiliation{Department of Pulse Plasma Systems, Institute of Plasma Physics CAS, Prague, Czech Republic}
\author{J\'an Tungli}
\affiliation{Department of Physical Electronics, Faculty of Science, Masaryk University, Brno, Czech Republic}
\author{Milan \v{S}imek}
\affiliation{Department of Pulse Plasma Systems, Institute of Plasma Physics CAS, Prague, Czech Republic}
\author{Zden\v ek Bonaventura}
\affiliation{Department of Physical Electronics, Faculty of Science, Masaryk University, Brno, Czech Republic}

\title{Electron multiplication in nanovoids at the initial stage of nanosecond discharge in liquid water}
\date{\today}

\begin{abstract}
The process of electron multiplication through the 
bouncing-like accelerated motion of electrons inside 
nanovoids formed owing to external electric fields in bulk 
liquid water is investigated using Monte Carlo simulations
in Geant4-DNA. 
Our results show that the initial charge developed at the metal/liquid
interface can be multiplied and expanded along the direction of the external electric field on a picosecond timescale, owing to collision-free interiors of the nanoruptures.
Characteristic features of two different electron multiplication mechanisms are revealed and characterized. We find that electrons can be accelerated inside cylindrical nanoruptures while bouncing off the void/water interface.
Simulations predict geometric conditions leading to charge multiplication along the void, rather than electron capture or thermalization in bulk liquid.
Our results are consistent with the recent verification of the causal relation between electrostriction-induced perturbations in bulk liquid, and the subsequent formation of luminous filaments evidencing the presence of energetic electrons.
\end{abstract}

%\pacs{}
\maketitle

\emph{Introduction}.---The processes of electron generation and multiplication govern the basic 
physics of electrical discharges, including 
discharges in polar liquids such as water. In discharges 
produced by highly non-uniform and time-dependent electric 
fields (usually produced through high-voltage pulses of 
nanosecond duration applied to the pin-plane electrode 
geometry), ultrafast processes on metal/liquid water 
interfaces and in H-bonded network structure of bulk liquid water 
provide competing mechanisms that determine the dynamics of 
discharge initiation. Investigating principal mechanisms 
responsible for electron multiplication and acceleration in 
such non-homogeneous and highly collisional environments is 
extremely challenging and of fundamental importance. 
%%%%%%%%%%%%%%%%%%%%%%%%%%%%%%%%%%%%%%%%%%%%%%%%%%%%%%%%%%%%%%%%%%%%%%%
%%%%%%%%%%%%%%%%%%%%%%%%%%%%%%%%%%%%%%%%%%%%%%%%%%%%%%%%%%%%%%%%%%%%%%%
The main difficulties in studying such 
phenomena include the very complex behavior of water exposed to strong and rapidly 
changing electric fields and extremely short characteristic temporal (sub-nanosecond) and spatial (sub-micrometer) scale.  
Another poorly understood phenomenon is the 
origin and evolution of the charged species leading to discharge 
initiation. Considerable uncertainty remains about processes
leading to the multiplication of principal charged species in external electric field, including their spatiotemporal expansion and the formation of branched structures in bulk liquid water on millimeter scale.

It should be noted that owing to H$_2$O autoionization \cite{Geissler2121, Volkov2017}, even chemically pure liquid water contains a certain number of H$_3$O$^+$ (hydronium) and OH$^-$ (hydroxide) ions, which are responsible for minimal electrical conductivity of $\approx$ 6$\times$10$^{-2}$ $\mu$S/cm. Autoionization occurs because of the fluctuations of local electric fields acting between neighboring molecules that can drive proton transfer, leading to the formation 
of hydronium and hydroxide ions. The nascent ions either 
recombine within tens of femtoseconds or get separated 
by the Grotthuss mechanism \cite{cukierman2006tu, imoto2020can}. 
\citet{shneider2016liquid} estimate equilibrium concentration of OH$^-$ ions in pure water at room temperature and density nanopores formed due to the electrostriction at sub-ns timescales to be $\approx$ 60 and $\leq$ 10$^4$ per $\mu$m$^3$, respectively. In real experiments, deionized water with initial conductivity  between 0.5--1 $\mu$S/cm is typically used. This implies characteristic   OH$^-$/nanopore ratio of $\approx$ 6$\times$10$^{-2}$ and a reasonable probability that OH$^{-}$ occurs on the nanopore/liquid interface (considering the dynamic nature of the H$_2$O autoionization process).  The OH$^-$ ions occuring on the nanopore/liquid interface then provide a source of free electrons. The electron density can be estimated \cite{shneider2016liquid} as $n_{\rm e} \sim w(I_{\rm n}, E) $[OH$^-$]$\Delta t$, where $w(I_{\rm n}, E)$ is the rate of electron tunnelling from OH$^-$ as a function of $I_{\rm n}$ the electron affinity energy in negative ion and $E$ is the electric field. $\Delta t$ denotes the characteristic time (typically $<$ 1 ns). \\
Other mechanisms that might further boost the presence of initial electrons include, e.g., natural background ionization and field ionization of water molecules at very high electric fields \cite{saitta2012ab} occurring locally due to the electrode surface micro-asperities or positive ions impact on the gas/liquid interface \cite{atrazhev2012breakdown}.

Moreover, water molecules in contact with metallic surfaces are affected by interfacial electric fields, in addition to chemisorption and H-bonding. This leads to the formation of 
the double layer structure on the interface (metal surface--Helmholtz layer--diffused layer--bulk liquid)~\cite{Stuve2012}. For positive potentials,
H$_2$O tends to re-orient with its hydrogen atoms away 
from the surface, and the ionization mechanism based on 
tunneling and proton transfer (H$_2$O $\rightarrow$ H$_2$O$^+$+e$^-$;  
H$_2$O$^+$+H$_2$O$\rightarrow$ H$_3$O$^+$+OH) becomes effective for electric fields above several GV/m~\cite{Stuve2012}.

The processes that alone, or through their interplay may explain the multiplication of charged species are impact ionization in the liquid bulk, field-assisted electron emission (i.e., Zener 
tunneling) \cite{gomer1994field, aghdam2020multiphysics, von2020nanosecond}, and the acceleration of electrons through low-density regions and ruptures resulting from electrostrictive forces
\cite{ando2012homogeneous,shneider2016liquid,li2020towards}. Pure impact ionization seems to be dubious, as electrons in the liquid are unable to gain sufficient energy to ionize water molecules. This is because the energy delivered by the applied electric field dissipates efficiently because of the frequent 
scattering of electrons on water molecules
\cite{migus1987excess,long1990femtosecond, 
laenen2000novel, signorell2020electron}. Moreover, the action of Zener 
tunneling as the enabling process seems improbable since it requires very 
large electric fields and a large pressure in the 
liquid volume \cite{joshi2004microbubbles, qian2005microbubble}. Nevertheless, Zener tunneling is supposed to play a role \cite{von2020nanosecond} at the metal/liquid interface owing to the electric field enhancement 
(tens of GV/m) due to local metallic surface asperities.

Currently, the prevailing opinion is that the multiplication mechanism 
is associated with the appearance of nanoruptures or nanovoids in the bulk liquid, which occurs as a result of the ponderomotive electrostrictive forces induced by highly non-uniform time-dependent electric fields 
\cite{Starikovskiy_2013,shneider2013dielectric,seepersad2015anode,vsimek2020investigation}.
These voids then provide sufficient collision-free space
for acceleration of initial and secondary electrons to energies exceeding 
the ionization potential of water molecules.
The initial seed electrons are supposed to be provided by the detachment from OH$^-$ ions present at the moment of void formation at the void/water interface. Some theories even assume that hydroxide density is increased at the cathode-side end of voids, which can further increase the injection of initial electrons into the voids \cite{zhang2020electron}.

Recent theoretical studies of nanosecond discharge in polar liquids based on electrostriction usually assume that cavities are 
initiated from fluctuations determined by the very fast switching dynamics 
of H-bonded H$_2$O molecules. Conditions under which these fluctuations can grow have been investigated assuming that spherical symmetry accounts for cavity expansion dynamics. The ability of electrons to ionize water molecules 
is then investigated based on the electron energy, after the diameter of the cavity is traversed.
It is important to stress that the spherical geometry of the cavities 
in external electric fields is a rather oversimplified approximation.
The strong electric field
is not only responsible for the expansion of the fluctuations 
but also causes rapid stretching in the direction of the electric field.
As a result, voids in the form of long fibrous hollow structures are created
in the bulk of water
\cite{shneider2013dielectric,kupershtokh2014three}.

\emph{Model and methods}.---Up to now, theoretical studies mostly considered the spherical cavity geometry and the one step acceleration-ionization mechanism.  
In this work, we propose a new scenario for electron
multiplication that can occur inside significantly stretched voids.
We show that electrons can propagate inside the cavity while bouncing off the water surface. This bouncing-like motion allows for the repeated acceleration of electrons and, thus, enhances the production of secondary electrons in the cavity. 
We study this phenomenon through Monte Carlo simulation,
and we formulate the minimal conditions that need to be fulfilled for electrons to multiply. To the best of our knowledge, such a study has not been tackled before, and the presented results shed light on one of the missing steps in electrostriction-based theories for fast discharge initiation in liquid water.

In the present model we consider cavities as 
long cylindrical voids of radius $R$ with a homogeneous electric field of strength $E$ oriented along the axis.
Note that the electron motion in the cylinder is analogous to projectile motion in a vertical tube subjected to the action of
the gravitational force. It can be shown that the product of 
the electric field strength and the cavity radius,
$E\cdot R$, is a scaling parameter for discussion of 
simulation results in different cylinder radii and electric field strengths.

We suppose that electrons in the void accelerate freely, i.e., without collisions, 
and interact only with water molecules when they penetrate the water surrounding the void.
This assumption is justified because the characteristic spatial scale of these voids is in the order 
of $100\,$nm, i.e., 
much shorter than the mean free path of electrons, which is approximately $6\,\mu$m in the corresponding equilibrium water vapor pressure.

\begin{figure}[h!]
\centering
\includegraphics[width=.99\columnwidth]{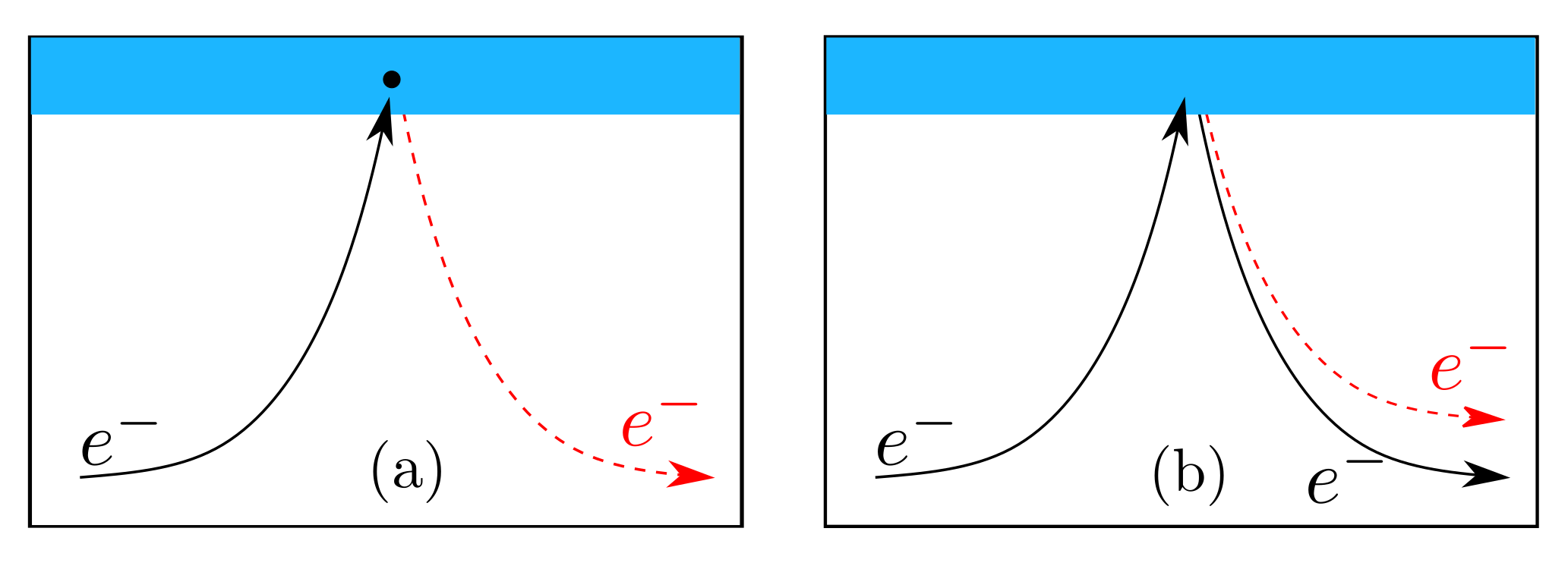}
        \caption{Electron interactions at the void/water interface: 
	(a) Termination: The primary electron thermalizes in the liquid and terminates there
	(b) Bouncing: the primary electron is bounced back to the void 
as a result of collisions with water molecules. 
New electrons can be emitted to the void during both processes (red dashed arrows).
\label{fig:ScenarioStep}
}
\end{figure}

The electron interaction with water near the surface of the void implies two basic outcomes: 
(a) the electron penetrates the water bulk and terminates there, or
(b) the electron is bounced back to the void and is accelerated by the electric field again.
These situations are schematically shown in Figure \ref{fig:ScenarioStep}.

%%%%%%%%%%%%%%%%%%%%%%%%%%%%%%%%%%%%%%
%%%%%%%%%%%%%%%%%%%%%%%%%%%%%%%%%%%%

In both cases, a certain number of secondary electrons may be emitted from the surface of water to the void. 
In the case of an electron penetrating the void/water 
interface, the electron's energy may dissipate through inelastic collisions. Ionizing 
collisions in the bulk liquid will produce H$_2$O$^+$ ions and 
secondary electrons. The H$_2$O$^+$ ions are converted to H$_3$O$^+$+OH, while electrons are hydrated in picosecond time scales \cite{Garret2005}. Excitation reactions driven by the low-energy electrons ($\leqslant$$10\,$eV) will produce H$_2$O$^{*}$ (where {}$^*$ stands 
for excited electronic states, e.g., ${\tilde {\rm a}}$$^3$B$^1$, $\tilde {\rm A}$$^{1}$B$^1$, 
${\tilde {\rm b}}$$^3$A$^1$, $\tilde {\rm B}$$^1$A$^1$,$\ldots$) 
\cite{Rescigno2013,matsui2016measuring}. 
All transient species formed during electron-driven 
reactions will be thermalized in the bulk liquid, forming 
a variety of species (e.g., H$_2$, H$_2$O$_2$ and OH$^-$)~\cite{Garret2005}.

In the following, we investigate conditions under which the
bouncing of electrons leads to an increase in the number of electrons propagating along the void. 
For the description of electron interactions with water,
we use the state-of-the-art simulation framework Geant4-DNA 
\cite{incerti2018geant4, bernal2015track, incerti2010comparison, incerti2010geant4}.
This framework offers a variety of models to simulate the physical interactions of electrons in liquid water. We use the Geant4-DNA physics `option 4' constructor, which includes the Emfietzoglou-Kyriakou
dielectric model for inelastic scattering and
the Uehara screened Rutherford model for elastic 
scattering of electrons \cite{kyriakou2015improvements}. The Sanche model is included for vibrational excitation \cite{Sanche2003},
and the Melton model for attachment \cite{Melton1972}. 
We assume that sub-excitation electrons, i.e., electrons with energy below $\varepsilon_{\rm se}$=7.4\,eV, 
cannot contribute to further ionization in water, as the acceleration of electrons due to the electric field is ineffective because of their high collisionality in water. 
Thus, no tracking of sub-excitation electrons immersed in water is performed in the simulation.
Note that this energy cutoff is a parameter of the simulation.

%%%%%%%%%%%%%%%%%%%%%%%%%%%%%%%%%%%%%%
\begin{figure*}[htb]
%\centering
\includegraphics[width=.99\textwidth]{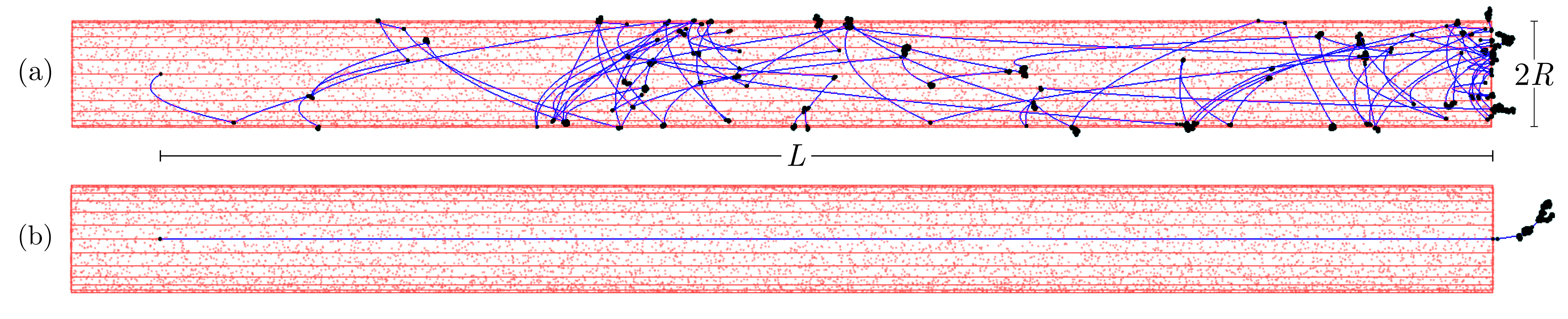}
        \caption{Two scenarios for electron propagation in long ruptures:
(a) bouncing and multiplication of electrons along the surface of the void,
(b) direct flight with ionization at the end of the rupture. Red lines and dots represent the profile of the cylindrical void, 
	blue lines show electron trajectories, and black dots describe locations of electron collision events.
	Geant4-DNA simulation was performed for $R$ = 30 nm, $E\cdot R = 25\,$V and length $L$ of the void $25R$ for a single initial electron with
	energy $7.4\,$eV.
	}
\label{fig:TwoCases}
\end{figure*}
%%%%%%%%%%%%%%%%%%%%%%%%%%%%%%%%%

%%%%%%%%%%%%%%%%%%%%%%%%%%%%%%%%%%%%%%
\begin{figure}[htb]
\centering
\includegraphics[width=.99\columnwidth]{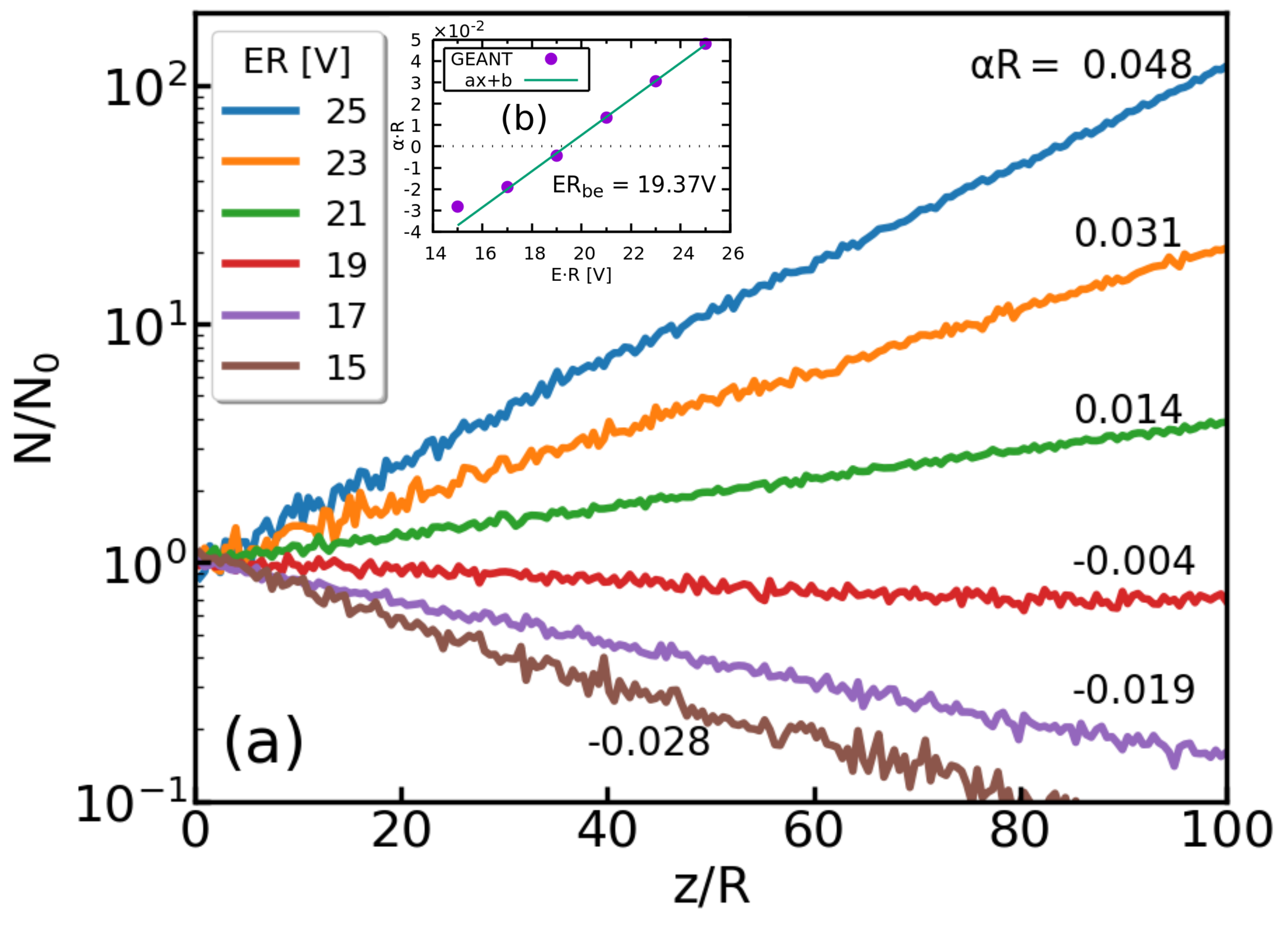}
        \caption{(a) Number of electrons propagating along 
the tube for $E \cdot R \in \{15,17,19,21,23, 25\}\,$V
	follows the exponential law 
$N(z/R)=N_0\exp\left(\alpha R z/R\right)$, where $N_0$ is the number of electrons at $z=0$, and $1/\alpha R$ 
is a characteristic e-fold multiplication distance.
(b) The coefficient $\alpha R$ as a function of $E \cdot R$. 
For $E \cdot R> 17\,$V, $\alpha R$ is a linear function
of $E \cdot R$. The value of $E \cdot R$ required for 
the conservation of electrons propagating along the tube is  
19.4 V,
i.e., when $\alpha R=0$.}
\label{fig:Avalanche}
\end{figure}
%%%%%%%%%%%%%%%%%%%%%%%%%%%%%%%%%

\begin{figure}[htb]
\centering
\includegraphics[width=.99\columnwidth]{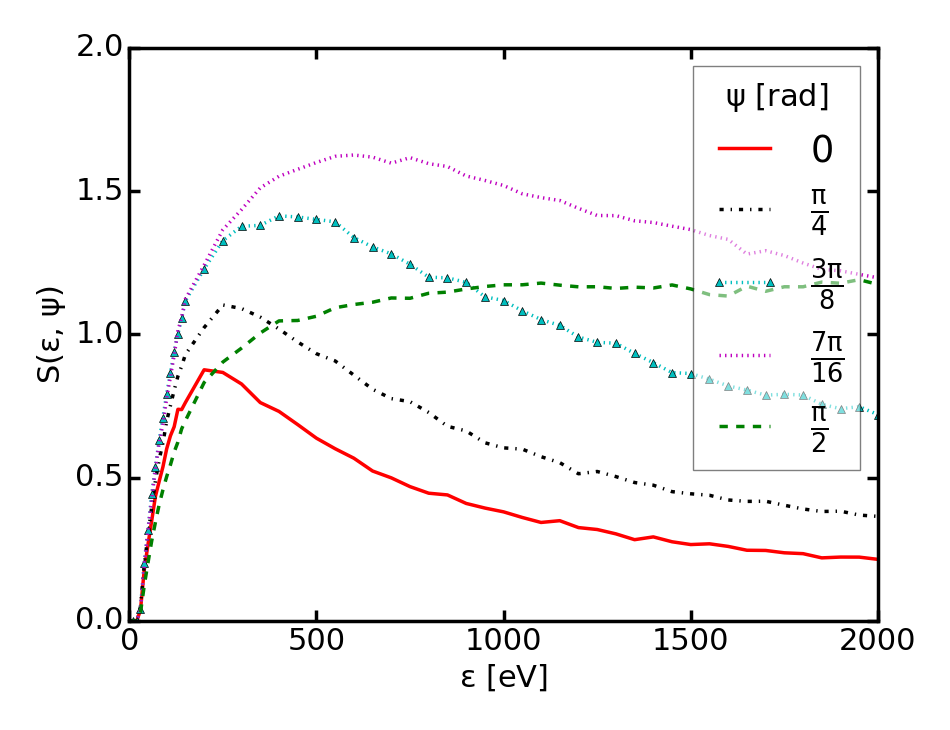}
\caption{Number of secondary electrons emitted from water by primary electron impact as a function of incident electron energy for five incident angles $\psi \in [0,\pi/2)$ calculated using Geant4-DNA.}
\label{fig:SEYs}
\end{figure}

\emph{Results}.---We compare two distinct scenarios for electron propagation
along the void in terms of the efficiency of electron multiplication.
The first scenario, shown in Figure \ref{fig:TwoCases}(a), considers bouncing of electrons with their possible multiplication along and at the end of the void. 
The second scenario, shown in Figure \ref{fig:TwoCases}(b), considers
a direct flight through the void with ionization possible only at the end of the void.
Concerning the model parameters, we choose $E \cdot R$ to vary between 15--25 V, which includes a break-even point between non-multiplying and multiplying cases. For long cylinders (e.g., $R$ = 30 nm), it corresponds to the homogeneous electric field varied between 0.5 GV/m and 0.83 GV/m. \\
Let us first focus on the scenario with bouncing-like propagation of electrons. 
In the Geant4-DNA simulation, an ensemble of primary electrons
is launched from the surface of water to the void with an isotropic velocity distribution at initial energy $7.4\,$eV. Despite this initial condition being somewhat arbitrary, there is no significance of it for the later propagation of electrons down the cavity.
This is because any memory of the initial condition rapidly disappears
once the electrons enter the bulk water. For the majority of initial electrons, this happens on the axial distance along the void, which is comparable to the diameter of the cylinder.\\
Figure \ref{fig:Avalanche}(a) shows the number of electrons 
propagating along the tube for $E \cdot R \in \{15, 17, 19, 21, 23, 25\}\,$V.
Note that the stationary growth or the decay in the number of electrons
can be characterized by the exponential law $N(z/R)=N_0\exp\left(R\alpha z/R\right)$, where $N_0$ is the number of electrons at $z=0$, and $1/(\alpha R)$ is a characteristic e-fold multiplication distance.
The transient phase of electron propagation with respect to the 
initial condition is not shown here and 
the length of the void is assumed to be sufficiently long ($L=150R$)
to ensure  that
the interaction of electrons with the end of the  void has no impact 
on observed results.
The parameter $\alpha R$ shows a linear dependence on $E \cdot R$ 
for $E \cdot R > 17\,$V, as seen in Figure~\ref{fig:Avalanche}(b). 
The break-even point value of $E \cdot R = 19.4\,$V for the electron number propagating along the tube is obtained from the condition $\alpha R = 0$.
For $E \cdot R > 19.4\,$V, we then observe an avalanche-like multiplication of electrons.
This avalanche is fed by the secondary emission of electrons from the void/water interface.
The number of electrons emitted to the void versus
the electron impact on the void/water interface is plotted in Figure \ref{fig:SEYs}
as a function of the incident electron energy $\varepsilon$ and the angle of incidence $\psi \in [0,\pi/2)$.
The number of electrons emitted first increases with the energy 
but then starts to decrease again for higher energies. This behavior
varies strongly with the impact angle. 
Nevertheless, the energy of the primary electron 
must be over 120 eV even for the most favorable impact angles
to ensure that the average number of emitted electrons per impact is higher than one. 

It is important to mention here that the number of electrons emitted depends
on the energy cutoff for the sub-excitation electrons: lower values for the cutoff would lead to higher electron emission. This would result in
a lower value of $E \cdot R$ required for the electron avalanche to be developed inside the cavity. Therefore, the value of $19.4\,$V should be 
regarded as an upper estimation of $E \cdot R$ that guarantees the occurrence of an avalanche. 
%%%%%%%%%%%%%%%%%%%%%%%%%%%%%%%%%%%%%
%%%%%%%%%%%%%%%%%%%%%%%%%%%%%%%%%%%%%

\begin{figure}[htb]
    \centering
    \includegraphics[width=0.99\columnwidth]{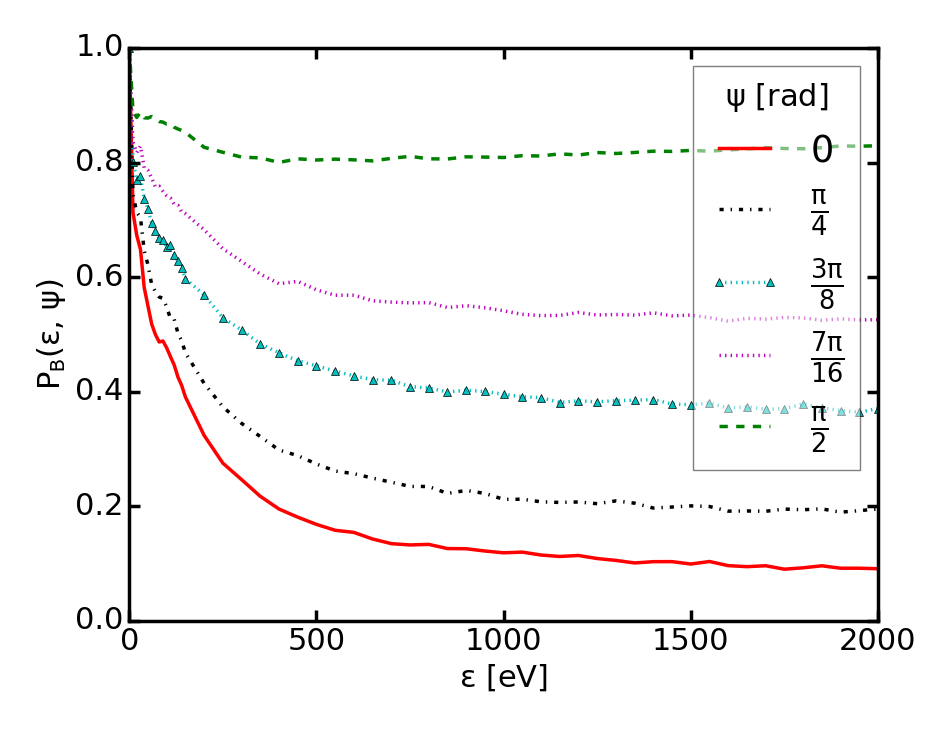}
        \caption{Probability of the primary electron bouncing 
back to the void $P_{\rm B} (\varepsilon, \psi)$ as a function of
the electron energy for five incident angles $\psi \in [0,\pi/2)$
calculated using the Geant4-DNA.}
    \label{fig:bouncing}
\end{figure}

As discussed above, the electrons that impact the void/water interface will
either terminate in water or bounce back to the void. 
The probability of the electron bouncing back to the 
void as a function of the incident electron energy 
and angle of incidence $\psi \in [0,\pi/2)$ is shown in Figure \ref{fig:bouncing}.
%%%%%%%%%%%%%%%%%%%%%%%%%%%%%%%%%%%%%
%%%%%%%%%%%%%%%%%%%%%%%%%%%%%%%%%%%%%
Clearly, this probability is lowest when the electron impacts the surface of water perpendicularly, and is the highest for grazing incidence. 
In addition, the bouncing probability decreases significantly with 
increase in the initial electron energy. This is due to the fact
that electrons with higher incident energies are able to penetrate deeper
beneath the void/water interface and, thus, it is less probable
that subsequent collisions will kick these electrons back to the void. 
Note that the probability of electron bouncing is complementary to the 
probability that the electron will terminate in bulk water.
%%%%%%%%%%%%%%%%%%%%%%%%%%%%%%%%%%%%%

\begin{figure}[htb]
    \centering
    \includegraphics[width=.99\columnwidth]{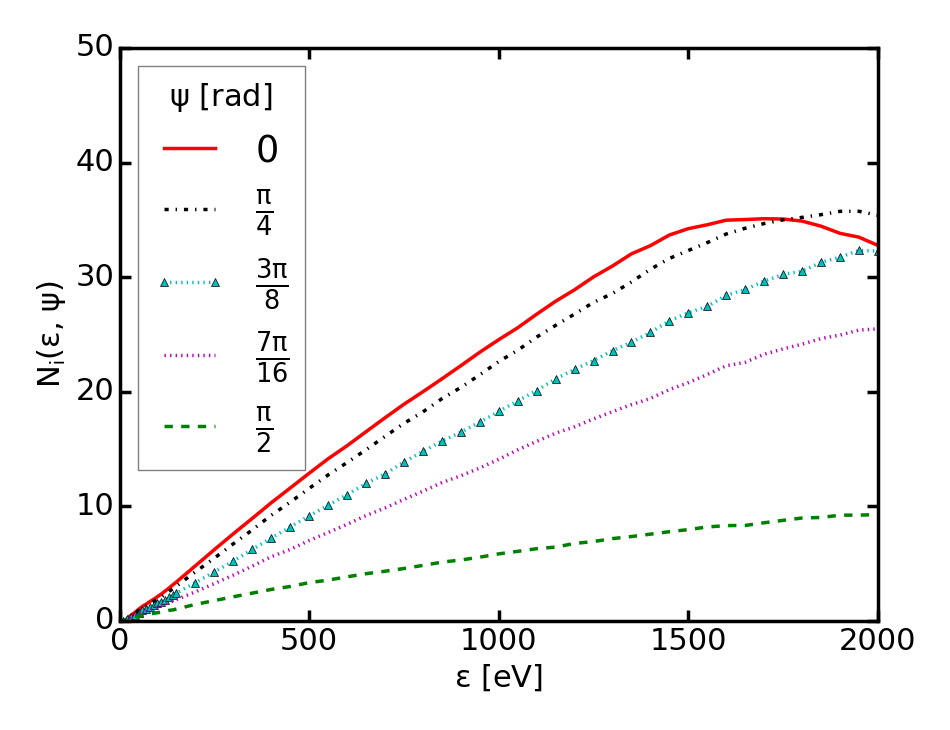}
        \caption{Number of secondary electrons generated by the impact of
primary electrons inside the liquid,
as a function of the electron energy for five incident angles $\psi \in [0,\pi/2)$ calculated using Geant4-DNA. \label{fig:ioni}}
\end{figure}

Another important quantity that describes interactions of
the primary electron with water is the number of secondaries that
are created in the bulk of water and thermalize inside.
This quantity is shown in Figure~\ref{fig:ioni}, and it increases with
increasing energy of the incident electron. 
Note that it decreases with the incident angle because of the increasing bouncing probability, as shown in Figure \ref{fig:bouncing}.

\begin{figure}[htb]
%\centering
\includegraphics[width=.99\columnwidth]{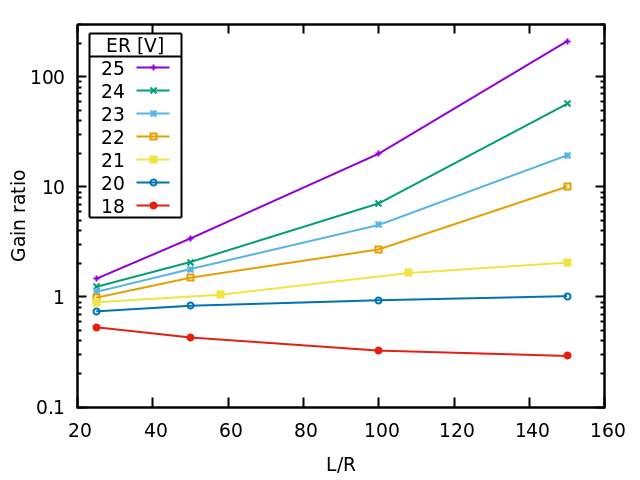}
        \caption{Ionization gain ratio: ${\cal N}_{\rm bp}/ {\cal N}_{\rm df}$, where ${\cal N}_{\rm bp}$ 
is the total number of ionizations for bouncing-like propagation, and ${\cal N}_{\rm df}$ is
the total number of ionizations for the direct flight, for 
$E \cdot R \in \{18, 20, 21, 22, 23, 24, 25\}\,$V as a function of the cavity length $L/R$.
\label{fig:GainRatio}}
\end{figure}

In the following, we compare the electron bouncing-like propagation scenario
with the direct flight scenario in terms of the total 
number of ionizing events generated in bulk water.
We denote the total number of ionizations recorded 
in case of bouncing-like propagation as ${\cal N}_{\rm bp}$.
Note that this number is a sum of all ionizations that were caused by
impacting electrons on the void/water interface,
both along the cavity surface, and at the end of the cavity. 
For the direct flight scenario, the same number of initial electrons 
at energy $\varepsilon_{\rm se}$ is launched
with velocity oriented directly along the axis of the tube. 
These electrons are accelerated by the applied electric field, 
and penetrate into the bulk of water at the end of the tube. 
All ionization events are recorded, and we denote this number as ${\cal N}_{\rm df}$. 
Figure~\ref{fig:GainRatio} shows the ionization gain ratio ${\cal N}_{\rm bp}/ {\cal N}_{\rm df}$
for $E~\cdot~R~\in~\{18,20,21,22,23,24,25\}\,$V, as a function of the cavity length $L/R$.
For $E \cdot R$ higher than the break-even point value, the bouncing-like propagation of electrons along the cavity is more efficient in terms of the total number of ionizations produced in bulk
water. Note that this efficiency is higher for higher values of $E \cdot R$.

Upon analyzing the various simulation results, we found that 
the bouncing-like propagation scenario can be characterized by a certain effective
propagation velocity that is in the order of $10^6\,$m/s. 
Considering particular conditions illustrated in Figure~\ref{fig:TwoCases}(a) 
(for $R=30\,$nm and $L/R= 25$), this velocity reaches about $3\times10^6\,$m/s.

\emph{Conclusions}.---We analyzed two scenarios for electron 
propagation/multiplication in nanovoids created and stretched by the action of strong 
pulsed electric fields in liquid water. 
We show that an electron avalanche in the void can
be created. This avalanche is fed by 
the secondary electron production on the void/water interface
and is enhanced by the possibility of electrons bouncing off the cavity 
surface. The avalanche is able to grow if the
product of the cavity radius and the electric field in the cavity
is higher than 19.4~V.
Comparing the total number of ionizing events that occur in the bulk
water because of this electron avalanche to the total number of ionizations
when electrons directly fly through the cavity shows 
that the bouncing-like propagation ensures higher ionization yield
when the condition for electron avalanche growth is satisfied.
Note that with respect to the value of the cutoff energy for electron tracing, all conclusions are valid. However, lowering the energy cutoff would 
lead to the enhancement of secondary emission of electrons in the cavity.
As a result, the electron ionization gain would be even higher for the
bouncing-like propagation scenario.

%%%%%%%%%%%%%%%%%%%%%%%%%%%%%%%%%%%%%%
%%%%%%%%%%%%%%%%%%%%%%%%%%%%%%%%%
We observe that the effective velocity of bouncing-like propagation is in the order of $10^6\,$m/s,
and this should represent the upper limit for the velocity of expansion of luminous fronts driven by avalanche electrons.
It is important to note that this $10^6\,$m/s limit is in agreement with recent experimental results evidencing an approximately linear initial expansion of the luminous fronts 
on distances in the order of hundreds of micrometers, with a propagation velocity $\sim 2\times10^5\,$m/s \cite{vsimek2017luminous}.

Moreover, the results of this work are consistent with recent experimental observations on the causal relation between two coupled dark (non-luminous) and luminous discharge phases \cite{vsimek2020disentangling}.
Experimental results confirmed that the luminous phase (evidencing the presence of high-energy electrons) implies the prior occurrence of the dark non-luminous phase (evidencing bulk liquid perturbed by the electrostriction).
Observed delays between the two phases in the order of hundreds of picoseconds support the present concept based on the onset
and growth of electron avalanches inside nanoruptures aligned along the applied external electric field.
Further insights might be gained by analyzing the characteristic parameters of the avalanches (electron density
and energy distribution function) developing in longer voids (tens of micrometers).\\

\noindent{\bf Acknowledgments}: Funding from the Czech Science Foundation (GA\v{C}R, GA 18-04676S) is gratefully acknowledged. ZB also acknowledges support by the project LM2018097 funded by the Ministry of Education, Youth and Sports of the Czech Republic. We~would like to thank Dr. Václav Štěpán (Department of Radiation Dosimetry, Nuclear Physics Institute of the Czech Academy of Sciences, Prague, Czech Republic) for his beneficial advice and the discussion concerning Geant4-DNA. We~would also like to thank Editage (www.editage.com) for English language editing.

\bibliography{biblio_article}

\end{document}